# A Scalable Feature Selection and Opinion Miner Using Whale Optimization Algorithm

Amir Javadpour[1], Samira Rezaei[2], Kuan-Ching Li[3] and Guojun Wang[0000-0001-9875-4182] 1(✉)

[1] School of Computer Science, Guangzhou University, Guangzhou, China, 510006
[2] Bernoulli Institute for Mathematics and Computer Science, University of Groningen, The Netherlands
[3] Department of Computer Science and Information Engineering, Providence University, Taiwan
Email: a_javadpour@e.gzhu.edu.cn, csgjwang@gzhu.edu.cn, kuancli@gm.pu.edu.tw

**ABSTRACT.** Due to the fast-growing volume of text document and reviews in recent years, current analyzing techniques are not competent enough to meet the users' needs. Using feature selection techniques not only support to understand data better but also lead to higher speed and also accuracy. In this article, the Whale Optimization algorithm is considered and applied to the search for the optimum subset of features. As known, F-measure is a metric based on precision and recall that is very popular in comparing classifiers. For the evaluation and comparison of the experimental results, PART, random tree, random forest, and RBF network classification algorithms have been applied to the different number of features. Experimental results show that the random forest has the best accuracy on 500 features.

**Keywords:** feature selection, Whale Optimization algorithm, selecting optimal, classification algorithm

## 1    INTRODUCTION

Nowadays, with the rapid growth of networks and social media, people can publish their ideas and thoughts directly on the internet. Reading and accessing to these opinions and comments on a variety of topics have been tremendously appealing [1]. On the other hand, a massive amount of generated data is not easy to handle. Users can not manually analysis these data and some summarization and classification algorithms are needed to reduce the entire data into meaningful classes. This is very important for cloud service providers as they face memory such as Cloud Computing-based Data Storage and Disaster Recovery [2]–[4]. Besides, as there are many contradictory ideas about a topic, one user needs to understand the positive and negative views and the reason of this contradiction in the available reviews [5]–[8]. This is where automatic analysis of comments and summarizing them into valuable information emerges and plays an vital role to generate scalable solutions[9]–[11].

The business world is also very dependent on this strategy as the customers prefer to read and analyze the previous customers' reviews before purchasing an item. A lot of online marketers have been created based on this need which customers want to be aware of other people's opinions about such products. Therefore, it is evident that the way to analyze and explore the ideas and beliefs of users can be useful for applications. Emotions can be defined as positive or negative feelings. Exploring the ideas is a computational technique for extracting, classifying, understanding, and obtaining the ideas expressed in various contents [11].

Karamatsis presented a system study that applies more than five dictionaries to classify comments in social into two categories [12]. Each dictionary has seven features for each message that was later used as an entry for SVM classifier. They tested their system with several datasets containing such as twits and comments on twitter. Authors in [13] used supervised statistical systems to design an emotion analyzer in the long term. The dictionary automatically generates an optimal set of features based on user's twits over a long period of time. Latent Dirichlet Allocation (LDA) is also used to cluster the document words into unsupervised learned topics using Poisson distribution in [14]. Multinomial and Dirichlet distributions are other options to use specially when the number of words is very high. For each word in the set of opinions, a specific topic is randomly selected by using the multinomial distribution.

Classifying emotions in medical domain using ant colony optimization algorithm (ACO) is done in [12]. The proposed classification framework has considered the content of medical documents which impart patients health information such as health status, obtained observations and examination results. They even might have evidence about diagnoses and interventions. Furthermore, analyzing and evaluating these data should be done wisely as it will be used to judge clinical outcomes or asses the impact of a medical condition on patient's wellbeing. In this study feature sets are extracted from reviews using Term Frequency-Inverse Document Frequency (TF-IDF) and then ACO has been applied. At the final step, they applied classification algorithms such Naïve Bayes, and Support Vector Machine (SVM) to the selected features.

Besides ACO, other metaheuristic algorithms also have been widely used for feature selection in opinion mining and sentiment analysis. Authors in [13] used firefly algorithm to optimize the exploration of comments in stock market. A hybrid model of metaheuristic algorithms of Whale Optimization Algorithm (WOA) and Simulated Annealing (SA) algorithm has been used in [15]. The main idea of using SA is to improve the output of WOA by exploring the most promising regions detected by WOA algorithm. Authors evaluated their method on 18 standard benchmark datasets of UCI repository and compared with three well-known wrapper feature selection methods in the literature. Other approaches such as Genetic Algorithm and Rough Set Theory [16] has also been used to extract and identify sentiment patterns on opinions by consumers. Authors compared the quality of traditional feature selection techniques with sentiment analysis methods by applying text classification.

Researchers in [17] analyzed the spam comments which artificially either promote or devalue the quality of a specific product or service. This study is based on the fact that many of features are redundant, irrelevant and noisy. Furthermore, extracting meaningful features from in a very high dimensional feature space has a crucial role in detecting spam comments. The extracted features by a hybrid model of Cuckoo and Harmony search are used to classify comments into spam and non-spam reviews using the Naive Bayes algorithm. Another study proposed sentiment mining to analyze customers' comments to handle the unstructured format of online reviews. Researchers considered the hybrid role of typed dependency relations (TDR) and part-of-speech tagging (POST) to find existing relation between features and sentiment words. These techniques also have been used to create a list of rules in identifying relations between features [18].

TABLE I. AN EXAMPLE OF PRESELECTED WORDS IN DIFFERENT CLASSES

| Class | A few examples of Preselected words |
|---|---|
| **Positive** | Easy, good, economic, Good, quality, inexpensive, Awesome, fun |
| **Negative** | eat, no, gulps, awkward, consumption, small, crazy, expensive |
| **Neutral** | sometimes, be, see Viewfinder, life, is, are, I, discover, media |

The rest of this article contains the following parts. In the next section, we review the proposed method and introduce the proposed feature selection algorithm in details. Then we present the result and the discussion. We will also examine the classification algorithms, the results of their application on the optimal set of features, and results of applying them on an optimal set of features. Finally, conclusions and future suggestions are explained.

1. MODELING AND PROBLEM SOLVING

In this section, we explain the details of the proposed method. Figure 1 illustrates the main steps of our method. By importing data into our system, we should take care of any incompatibility of data and our programing and simulating software. To be able to handle texts in Matlab and apply feature selection, we need to operate data preprocessing. The software that we have used to apply data preprocessing on is Visual Studio, and the selected programming language is C#. We defined three classes to put each word in the sentences on. The defined classes are positive words, negative and neutral words. In the preprocessing step we also made some changes based on this word division. The idea behind this is to create a dataset of positive and negative words for the next steps and select most significant words among them.

Regarding the effect of feature selection in achieving higher accuracy of classification algorithms, feature selection in this research has been utilized through the method of feature selection of the Whale Optimization algorithm. Once the top features are selected, we applied different classification algorithms such as random forest, RBF network, random tree, and PART. Finally, we compare the results based on the evaluation criteria. The database used in this research is available at the Illinois University Data Center[19] and has two sections in which the users' positive and negative views are included for a mobile product. The number of positive comments in the database is 22940, and the number of negative comments is 22935, so the total number of records in this study is 45875.

1. PREPARING OF DATA

Preprocessing is one of the essential steps in any machine learning algorithms. There are two text files containing negative and positive comments as our input data. We applied initial processing steps to be able to use this data for furture analysis. The methodology is based on the presented method in [20], in which the words in the comments are divided into three categories of words representing a positive opinion, the words representing negative opinions and neutral words. In order to do this, two sets of data containing positive words and negative words are extracted by reviewing the opinions in the dataset, as shown in Table I. After extracting positive and negative words, all records in the database are examined according to the positive and negative lists.

The process of achieving the optimal set of features is that according to a study [21] that explores the methods of feature selection in the users' comments. The number of repetitions per word in the entire data has been calculated, and the following criteria have been applied to emit some words from our positive, negative and neutral words.

    i.    The list of words which have been repeated only once or twice in the entire dataset.

    ii.    In order to remove prepositions, words with repetitions more than half the number of records are deleted from the vocabulary lists.

Therefore, in the selected initial set of words, words with a frequency less than two and more than half the total number of comments are removed from the data. Also, in order to prevent any mistakes in lowercase and uppercase letters, all letters are converted into lowercase. The total number of records containing the negative and positive comments are 22935 and 22940 respectively. The entire preprocessing step is implemented in Visual Studio software with C# programming language.

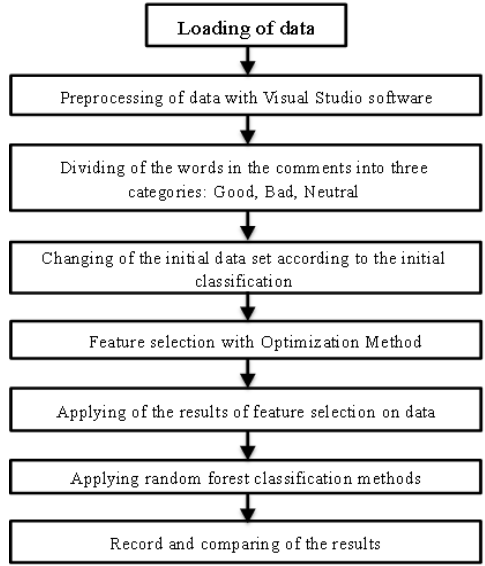

Fig. 1. Flowchart of Proposed Method.

2. **FEATURE SELECTION**

The Whale optimization algorithm is a relatively new meta-heuristic algorithm that imitates humpback whales hunting behavior. The significant difference between this algorithm and other heuristic algorithms is the stimulation of random hunting behavior. In other words, this algorithm is based on the behavior of humpback whales to identify the location of prey and encircle it. At each iteration, WOA algorithm assumes that the current best candidate is the optimum solution or is close it. Since the position of the optimal design in the search space is not known a priori, the WOA algorithm assumes that the current best candidate solution is the target prey or is close to the optimum. After the best search agent is defined, the other search agents will hence try to update their positions towards the best search candidate. Other search agents update their position based on the position of the best current solution [22]. Initially, the parameters related to the Whale optimization algorithm are set according to the conditions of the problem. In the next step, it is necessary that through fitting function, we evaluate the results achieved at each stage and improve the results. Table II shows the parameters used in this algorithm.

TABLE II.  PARAMETERS USED IN THE WHALE OPTIMIZATION ALGORITHM.

| Parameters | Description | Quantity |
| --- | --- | --- |
| SearchAgents_no | The number of search factors | 30 |
| MaxIt | Maximum number of rings | 100 |
| lb | lower limit | 0 |
| ub | Upper limit | 4000 |
| dim | Number of Dimensions | 100 |

a) **Fit function**

In order to select the most critical features for opinion mining, several methods can be used to compute the fit function. In this paper, we used the F-Measure's criteria to evaluate the results. It is a combination of two criteria of precision and recall that are extensively considered in calculating the quality of the classifications. The concepts of FP, TP, and FN in this paper are defined according to the textual data content as follows[3]:

TN: Indicates the number of records whose actual category is negative, and the proposed system could also correctly identify this category.

TP: Represents the number of records whose real category is positive, and the proposed algorithm also correctly recognizes this category.

FP: Represents the number of records that are negative in their actual data but the proposed algorithm has mistakenly identified it as positive

FN: Indicates the number of records whose actual data is positive, but the proposed algorithm has put this record in a negative category.

$$Precision = \frac{TP}{TP+FP} \quad (1)$$

$$Recall = \frac{TP}{TP+FN} \quad (2)$$

$$Accuracy = \frac{TP}{TP + FP + TN + FN} \quad (3)$$

$$f(measure) = \frac{2*recall + precision}{recall + precision} \quad (4)$$

TABLE III. EXAMINING THE VARIATIONS OF 10 WORDS IN THE SELECTED COLLECTION OF WORDS.

| Repetitions | The word in the positive selection of words |
|---|---|
| The first repeat | 'forever.', 'outstanding', 'numbers', 'hardly', 'efficiency', 'clam', 'full', 'reception!, 'developing', 'add' |
| The tenth repeat | forever.', 'outstanding', 'numbers', 'low', 'efficiency', 'clam', 'batteries', 'reception!', 'developing', 'add' |
| The fiftieth repeat | forever.', 'outstanding', 'camcorder', 'easy-to-use', 'efficiency', 'beautiful', 'clam' 'batteries', 'reception!', 'developing' |

b) **Applying classification algorithms**

At this stage, considering the best wordlist from the feature selection step, we will apply a random forest algorithm. The implementation of this part of the simulation has been done with the Weka open-source tool. The random forest algorithm has a huge potential to become a popular method for future classifications since its performance is comparable to other group methods, including Bagging and Boosting. This algorithm, as a voting group algorithm, produces several different decision trees as the basic categories and applies the majority vote to the original tree results. In order to evaluate the proposed method, the k-fold method has been used.

3. **SIMULATION AND RESULTS**

With the implementation of the Whale optimization algorithm in the feature selection phase, in each repetition, the algorithm tries to find the optimum solution according to the fit function. At first, a collection of words is extracted randomly from positive and negative words, and the final result is the best vocabularies that can optimize the defined fit function. The results are reported in Table III. The number of selected features (selected words) is 100, in which 50 are positive, and 50 are related to the negative wordlist. The results of applying classification algorithms on the data, considering the set of selected features in the feature selection step have been followed up. By examining the results, it is observed that the random forest algorithm performs better than all other classifier algorithms in all of the evaluation criteria. However, the running time for this algorithm is higher compared to other methods.

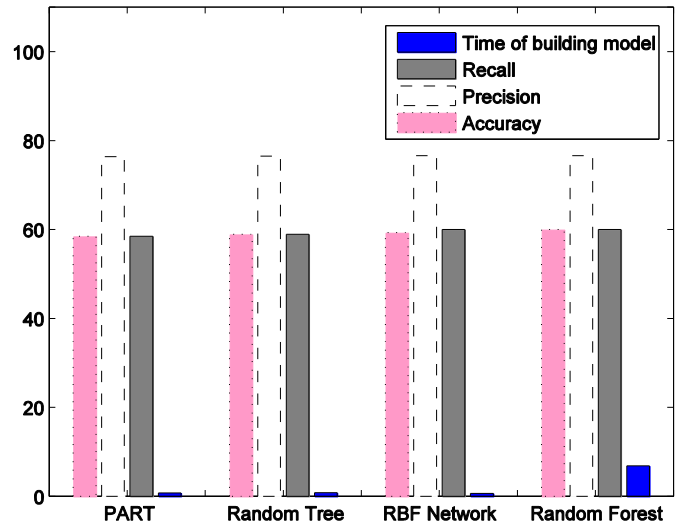

Fig. 2. Comparison of the quality indices obtained from applying classifier algorithms on 100 selected features.

### a) Change the number of features

In this section, we have changed the number of features selected in the Whale algorithm. The number of selected words in the Whale algorithm has been changed to 200 and 500. Due to the increasing the number of words, we expect the program's running time to increase. By comparing the obtained results, it is observed that with the increasing number of features, the accuracy of the fit function is much improved. This means that by increasing the number of features, we have been able to achieve better results from the set of features or words in our evaluation function. Table IV and Figure 4 shows the results of the applying classification algorithms on the selected data set with 200 and 500 words (features).

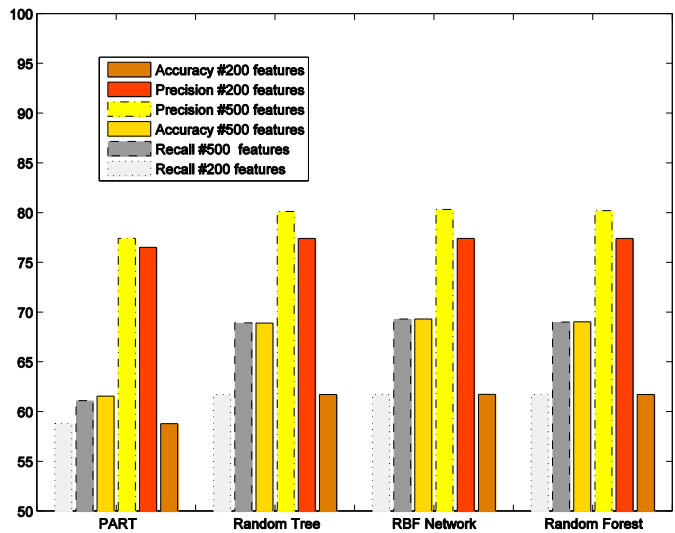

Fig. 3. chart of changing the value of the fit function in different repetitions.

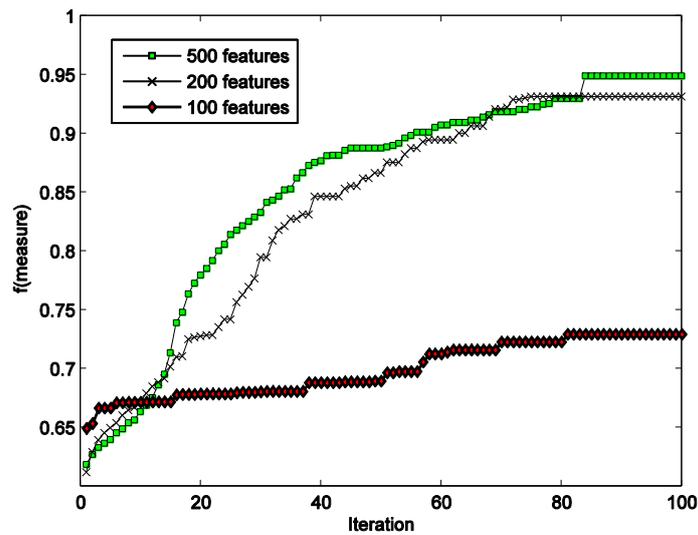

Fig. 4. Results of the applying of classification algorithms on the selected data.

### 4. CONCLUSION

Social media is the most crucial advertisement strategy for most Internet marketing companies. Although in the past advertising and public relations have played a significant role in firm brand promotion, it is social media based on virtual social networks that measure the value of organizations, companies, products, and services that have dominated in the last decades. In this research, the Whale optimization algorithm has been used to select the best set of features, aimed at reducing the dimensions of the data in the mining opinion or review classification. We tested our model over different number of features. We started with 100 features, and then we compared the classification quality by choosing 200 and 500 features. In this study, we applied different classification algorithms such as RBN network, random tree, and random forest. We recorded the accuracy of the obtained methods. We propose a hybrid model of metaheuristic algorithms for feature selection. Also, other fit functions such as information gain can be used to evaluate the obtained results.

TABLE IV. EVALUATION CRITERIA DERIVED FROM THE APPLIED METHODS FOR 100 FEATURES WITH K-FOLD EVALUATION.

| Evaluation criteria | PART | Random tree | RBF network | RF |
|---|---|---|---|---|
| **Accuracy** | 58.48 | 58.92 | 59.3 | 60 |
| **Precision** | 76.4 | 76.5 | 76.6 | 76.6 |
| **Recall** | 58.5 | 58.9 | 60 | 60 |
| **Time of building model** | 0.76 | 0.82 | 0.65 | 6.86 |

TABLE V. EVALUATION OF THE APPLIED METHODS WITH K-FOLD EVALUATION.

| criteria | # Features | PART | Random tree | RBF network | RF |
|---|---|---|---|---|---|
| **Accuracy** | 200 | 58.78 | 61.7 | 61.728 | 61.71 |
| **Accuracy** | 500 | 61.55 | 68.89 | 69.297 | 69.02 |
| **Precision** | 200 | 76.5 | 77.4 | 77.4 | 77.4 |
| **Precision** | 500 | 77.4 | 80.1 | 80.3 | 80.2 |
| **Recall** | 200 | 58.8 | 61.7 | 61.7 | 61.7 |
| **Recall** | 500 | 61.1 | 68.9 | 69.3 | 69 |

**ACKNOWLEDGMENT**

This work was supported in part by the National Natural Science Foundation of China under Grant 61632009, in part by the Guangdong Provincial Natural Science Foundation under Grant 2017A030308006, and in part by the High-Level Talents Program of Higher Education in Guangdong Province under Grant 2016ZJ01.

**REFERENCES**

[1] J. Fresneda and D. Gefen, "Automatic Methods for Online Review Classification: An Empirical Investigation of Review Usefulness---An Abstract," in *Creating Marketing Magic and Innovative Future Marketing Trends: Proceedings of the 2016 Academy of Marketing Science (AMS) Annual Conference*, M. Stieler, Ed. Cham: Springer International Publishing, 2017, pp. 1331–1332.

[2] Z. Jian-hua and Z. Nan, "Cloud Computing-based Data Storage and Disaster Recovery," in *2011 International Conference on Future Computer Science and Education*, 2011, pp. 629–632.

[3] A. Javadpour, S. Kazemi Abharian, and G. Wang, "Feature Selection and Intrusion Detection in Cloud Environment Based on Machine Learning Algorithms," *2017 IEEE Int. Symp. Parallel*

| | |
|---|---|
| | *Distrib. Process. with Appl. 2017 IEEE Int. Conf. Ubiquitous Comput. Commun.*, pp. 1417–1421, 2017. |
| [4] | A. Javadpour, G. Wang, S. Rezaei, and S. Chend, "Power Curtailment in Cloud Environment Utilising Load Balancing Machine Allocation," in *2018 IEEE SmartWorld, Ubiquitous Intelligence Computing, Advanced Trusted Computing, Scalable Computing Communications, Cloud Big Data Computing, Internet of People and Smart City Innovation (SmartWorld/SCALCOM/UIC/ATC/CBDCom/IOP/SCI)*, 2018, pp. 1364–1370. |
| [5] | W. Yang, G. Wang, K.-K. R. Choo, and S. Chen, "HEPart: A balanced hypergraph partitioning algorithm for big data applications," *Futur. Gener. Comput. Syst.*, 2018. |
| [6] | Q. Zhang, Q. Liu, and G. Wang, "PRMS: A Personalized Mobile Search Over Encrypted Outsourced Data," *IEEE Access*, vol. 6, pp. 31541–31552, 2018. |
| [7] | S. Zhang, G. Wang, M. Z. A. Bhuiyan, and Q. Liu, "A Dual Privacy Preserving Scheme in Continuous Location-Based Services," *IEEE Internet Things J.*, p. 1, 2018. |
| [8] | F. Ali, K.-S. Kwak, and Y.-G. Kim, "Opinion mining based on fuzzy domain ontology and Support Vector Machine: A proposal to automate online review classification," *Appl. Soft Comput.*, vol. 47, pp. 235–250, 2016. |
| [9] | A. Javadpour and H. Memarzadeh-Tehran, "A wearable medical sensor for provisional healthcare," *ISPTS 2015 - 2nd Int. Symp. Phys. Technol. Sensors Dive Deep Into Sensors, Proc.*, pp. 293–296, 2015. |
| [10] | A. Javadpour, H. Memarzadeh-Tehran, and F. Saghafi, "A temperature monitoring system incorporating an array of precision wireless thermometers," *Smart Sensors and Application (ICSSA), 2015 International Conference on*. pp. 155–160, 2015. |
| [11] | J. A. Balazs and J. D. Velásquez, "Opinion mining and information fusion: a survey," *Inf. Fusion*, vol. 27, pp. 95–110, 2016. |
| [12] | R.-M. Karampatsis, J. Pavlopoulos, and P. Malakasiotis, "AUEB: Two Stage Sentiment Analysis of Social Network Messages," in *SemEval@COLING*, 2014. |
| [13] | X. Zhu, S. Kiritchenko, and S. Mohammad, "NRC-Canada-2014: Recent Improvements in the Sentiment Analysis of Tweets," in *Proceedings of the 8th International Workshop on Semantic Evaluation (SemEval 2014)*, 2014, pp. 443–447. |
| [14] | D. T. Santosh, K. S. Babu, S. D. V Prasad, and A. Vivekananda, "Opinion Mining of Online Product Reviews from Traditional LDA Topic Clusters using Feature Ontology Tree and Sentiwordnet," *Int. J. Educ. Manag. Eng.*, vol. 6, no. 6, p. 34, 2016. |
| [15] | M. M. Mafarja and S. Mirjalili, "Hybrid Whale Optimization Algorithm with simulated annealing for feature selection," *Neurocomputing*, vol. 260, pp. 302–312, 2017. |
| [16] | S. R. Ahmad, A. A. Bakar, and M. R. Yaakub, "Metaheuristic algorithms for feature selection in sentiment analysis," in *2015 Science and Information Conference (SAI)*, 2015, pp. 222–226. |
| [17] | S. P. Rajamohana, K. Umamaheswari, and S. V Keerthana, "An effective hybrid Cuckoo Search with Harmony search for review spam detection," in *2017 Third International Conference on Advances in Electrical, Electronics, Information, Communication and Bio-Informatics (AEEICB)*, 2017, pp. 524–527. |
| [18] | S. R. Ahmad, M. R. Yaakub, A. A. Bakar, and others, "Detecting Relationship between Features and Sentiment Words using Hybrid of Typed Dependency Relations Layer and POS Tagging (TDR Layer POS Tags) Algorithm," *Int. J. Adv. Sci. Eng. Inf. Technol.*, vol. 6, no. 6, pp. 1120–1126, 2016. |
| [19] | The University of Illinois at Chicago, "Opinion Mining, Sentiment Analysis, and Opinion Spam Detection," *https://www.cs.uic.edu/~liub/FBS/sentiment-analysis.html*, 2017. . |
| [20] | K. Zhang, H. Xu, J. Tang, and J. Li, "Keyword Extraction Using Support Vector Machine," in *Advances in Web-Age Information Management: 7th International Conference, WAIM 2006, Hong Kong, China, June 17-19, 2006. Proceedings*, J. X. Yu, M. Kitsuregawa, and H. V. Leong, Eds. Berlin, Heidelberg: Springer Berlin Heidelberg, 2006, pp. 85–96. |
| [21] | G. Forman, "An extensive empirical study of feature selection metrics for text classification," *J. Mach. Learn. Res.*, vol. 3, no. Mar, pp. 1289–1305, 2003. |
| [22] | S. Mirjalili and A. Lewis, "The whale optimization algorithm," *Adv. Eng. Softw.*, vol. 95, pp. 51–67, 2016. |